\input{aipcheck}

\documentclass[
    ,final            % use final for the camera ready runs
  ]
  {aipproc}

\layoutstyle{6x9}
\usepackage{amsmath}
\begin{document}

%\small{

\title{An Overview of Neutrino Masses and Mixing in SO(10) Models}

\author{Mu-Chun Chen}{
  address={HET Group, Physics Department, Brookhaven National Laboratory,
Upton, NY 11973}
}

\author{K.T. Mahanthappa}{
  address={Department of Physics, University of Colorado, Boulder, CO 80309}
}

\begin{abstract}
We review in this talk various SUSY SO(10) models. Specifically, we discuss 
how small neutrino masses are generated in and generic predictions of 
different $SO(10)$ models. A comparison of the predictions 
of these models for $\sin^{2}\theta_{13}$ is given.
\end{abstract}

\maketitle

%%%%%%%%%%%%%%%%%%%%%%%%%%%%%%%%%%%%%%%%%%%%
%% MAINMATTER
%%%%%%%%%%%%%%%%%%%%%%%%%%%%%%%%%%%%%%%%%%%%

The flavor problem with hierarchical fermion masses and mixing has attracted a
great deal of attention especially since the advent of the atmospheric
neutrino oscillation data from Super-Kamiokande indicating
non-zero neutrino masses. The non-zero neutrino masses give support to the
idea of grand unification based on $SO(10)$ in which all the 16 fermions
(including the right-handed neutrinos) can be accommodated in one single spinor
representation. Furthermore, it provides a framework in which seesaw
mechanism arises naturally. Models based on $SO(10)$ combined with a 
continuous or discrete flavor symmetry group have been constructed 
to understand the flavor problem, especially the small neutrino masses and 
the large leptonic mixing angles. These models can be classified 
according to the family symmetry implemented in the model. 
We review in this talk how small masses and large mixing angles in the 
neutrino sector are generated in SO(10) models, and the unique 
predictions of each class of models. We also discuss other mechanisms that have 
been proposed to solve the problem of neutrino masses and mixing. For a more 
exhaustive list of references and detailed discussion, we refer the readers 
to our recent review\cite{Chen:2003zv} on which this talk is based.

{\bf Symmetric textures:} This type of models have been considered, for example, 
in Ref.[2-4]. SO(10) breaks down through 
the left-right symmetry breaking chain, which ensures the mass matrices are symmetric.
The Higgs content of this type of models contains fields in 10, 45, 54, 126 representations,
with 10, 126 breaking EW symmetry and generating fermions masses, and 
45, 54, 126 breaking SO(10). 
The mass hierarchy can arise if there is an $SU(2)_{H}$ symmetry acting non-trivially 
on the first two generations such that the first two generations transform as a doublet and 
the third generation transforms as a singlet under $SU(2)_{H}$, which breaks down at two steps, 
$\scriptstyle SU(2) {\epsilon M \atop \rightarrow} U(1) 
{\epsilon^{'} M \atop \rightarrow} nothing$ where $\scriptstyle \epsilon^{'} 
<< \epsilon << 1$. 
The mass hierarchy is generated 
by the Froggatt-Nielsen mechanism which requires the 
flavon fields acquiring VEV's along the directions 
specified in Ref.[2-4].
The resulting mass matrices at the GUT scale are given by
\begin{equation}\scriptstyle{
M_{u,\nu_{LR}}=
\left( \begin{array}{ccc}
\scriptstyle{0} & 
\scriptstyle{0} & 
\scriptstyle{\left<10_{2}^{+} \right> \epsilon'}\\
\scriptstyle{0} & 
\scriptstyle{\left<10_{4}^{+} \right> \epsilon} & 
\scriptstyle{\left<10_{3}^{+} \right> \epsilon} \\
\scriptstyle{\left<10_{2}^{+} \right> \epsilon'} & \
\scriptstyle{\left<10_{3}^{+} \right> \epsilon} &
\scriptstyle{\left<10_{1}^{+} \right>}
\end{array} \right)
= 
\left( \begin{array}{ccc}
\scriptstyle{0} & 
\scriptstyle{0} & 
\scriptstyle{r_{2} \epsilon'}\\
\scriptstyle{0} & 
\scriptstyle{r_{4} \epsilon} & 
\scriptstyle{\epsilon} \\
\scriptstyle{r_{2} \epsilon'} & 
\scriptstyle{\epsilon} & 
\scriptstyle{1}
\end{array} \right) M_{U}}
\end{equation}
\begin{equation}\scriptstyle{
M_{d,e}=
\left(\begin{array}{ccc}
\scriptstyle{0} & 
\scriptstyle{\left<10_{5}^{-} \right> \epsilon'} & 
\scriptstyle{0} \\
\scriptstyle{\left<10_{5}^{-} \right> \epsilon'} &  
\scriptstyle{(1,-3)\left<\overline{126}^{-} \right> \epsilon} & 
\scriptstyle{0}\\ 
\scriptstyle{0} & 
\scriptstyle{0} & 
\scriptstyle{\left<10_{1}^{-} \right>}
\end{array} \right)
=
\left(\begin{array}{ccc}
\scriptstyle{0} & 
\scriptstyle{\epsilon'} & 
\scriptstyle{0} \\
\scriptstyle{\epsilon'} &  
\scriptstyle{(1,-3) p \epsilon} & 
\scriptstyle{0}\\
\scriptstyle{0} & 
\scriptstyle{0} & 
\scriptstyle{1}
\end{array} \right) M_{D}}
\end{equation}
The right-handed neutrino mass matrix is of the same form as $\scriptstyle M_{\nu_{LR}}$ 
\begin{equation}\scriptstyle{
M_{\nu_{RR}}=  
\left( \begin{array}{ccc}
\scriptstyle{0} & 
\scriptstyle{0} & 
\scriptstyle{\left<\overline{126}_{2}^{'0} \right> \delta_{1}}\\
\scriptstyle{0} & 
\scriptstyle{\left<\overline{126}_{2}^{'0} \right> \delta_{2}} & 
\scriptstyle{\left<\overline{126}_{2}^{'0} \right> \delta_{3}} \\ 
\scriptstyle{\left<\overline{126}_{2}^{'0} \right> \delta_{1}} & 
\scriptstyle{\left<\overline{126}_{2}^{'0} \right> \delta_{3}} &
\scriptstyle{\left<\overline{126}_{1}^{'0} \right> }
\end{array} \right)
= 
\left( \begin{array}{ccc}
\scriptstyle{0} & 
\scriptstyle{0} & 
\scriptstyle{\delta_{1}}\\
\scriptstyle{0} & 
\scriptstyle{\delta_{2}} & 
\scriptstyle{\delta_{3}} \\ 
\scriptstyle{\delta_{1}} & 
\scriptstyle{\delta_{3}} & 
\scriptstyle{1}
\end{array} \right) M_{R}
\label{Mrr}}
\end{equation}
Note that, since we use
$\overline{126}$-dimensional representations of Higgses to generate the heavy
Majorana neutrino mass terms, R-parity is preserved at all energies. 
The effective neutrino mass matrix is 
\begin{equation}\scriptstyle{
\label{eq:Mll}
M_{\nu_{LL}}=M_{\nu_{LR}}^{T} M_{\nu_{RR}}^{-1} M_{\nu_{LR}}
= \left( 
\begin{array}{ccc}  
\scriptstyle{0} & \scriptstyle{0} & \scriptstyle{t} \\
\scriptstyle{0} & \scriptstyle{1} & \scriptstyle{1+t^{n}} \\  
\scriptstyle{t} & \scriptstyle{1+t^{n}} & \scriptstyle{1} 
\end{array} \right) \frac{d^{2}v_{u}^{2}}{M_{R}}} 
\end{equation}
giving rise to maximal mixing angle for the atmospheric neutrinos and LMA solution 
for the solar neutrinos. The form of the neutrino mass matrix in this model 
is invariant under the seesaws mechanism. 
The value of $U_{e3}$ is predicted to be large, 
close to the sensitivity of current experiments. This is a consequence of the solar angle 
being large. The prediction for the rate of $\mu \rightarrow e\gamma$ is about two orders 
of magnitude below the current experimental bound.

{\bf Lopsided/Asymmetric textures:} This type of models have been 
considered, for example, in  
Ref.[5-12]. In this case, $SO(10)$ breaks down to SM through the $SU(5)$ breaking chain. 
The Higgs sector of the model contains $10, 16, 45, 54$, with 
$\scriptstyle <16_{H_{1}}>$ breaking $SO(10)$ down to $SU(5)$ and 
$\scriptstyle <16_{H_{2}}>$ breaking the EW symmetry. 
The lopsided textures arise due to the operator $\scriptstyle 
\lambda (16_{i}16_{H_{1}})(16_{j}16_{H_{2}})$
which gives rise to mass terms for the charged leptons and down quarks, 
satisfying the SU(5) relation $\scriptstyle M_{d} = M_{e}^{T}$. 
When other operators are included, the lopsided structure of $M_{e}$ results, 
provided the coupling $\lambda$ is of order 1,
\begin{eqnarray}
\scriptstyle M_{u,\nu_{LR}} = 
\left(\begin{array}{ccc}
\scriptstyle \eta & \scriptstyle 0 & \scriptstyle 0\\
\scriptstyle 0 & \scriptstyle 0 & \scriptstyle (1/3,1) \epsilon\\
\scriptstyle 0 & \scriptstyle -(1/3,1)\epsilon & \scriptstyle 1
\end{array}\right) \scriptstyle \cdot m_{u}
\\
\scriptstyle M_{d} =  \scriptstyle \left(\begin{array}{ccc}
\scriptstyle \eta & \scriptstyle \delta & \scriptstyle \delta^{'}e^{i\phi}\\
\scriptstyle \delta & \scriptstyle 0 & \scriptstyle \lambda+\epsilon/3\\
\scriptstyle \delta^{'}e^{i\phi} & \scriptstyle -\epsilon/3 & \scriptstyle 1
\end{array}\right)\scriptstyle \cdot m_{d}
, \quad
\scriptstyle M_{e} =  \scriptstyle \left(\begin{array}{ccc}
\scriptstyle \eta & \scriptstyle \delta & \scriptstyle \delta^{'}e^{i\phi}\\
\scriptstyle \delta & \scriptstyle 0 & \scriptstyle -\epsilon\\
\scriptstyle \delta^{'}e^{i\phi} & \scriptstyle \lambda+\epsilon & \scriptstyle 1
\end{array} \scriptstyle \right)\cdot m_{d}.
\end{eqnarray}
The large mixing in $\scriptstyle U_{e,L}$ contributes to large leptonic mixing matrix, 
leading to the large atmospheric mixing angle. 
Meanwhile, because large mixing in $\scriptstyle U_{e,L}$ corresponds to large mixing in 
$\scriptstyle U_{d,R}$, the CKM mixing angles can be retained small. 
An unique prediction of the lop-sided models is the large branching ratio for LFV processes, 
e.g. $\mu \rightarrow e \gamma$. 
By considering neutrino RH Majorana mass term of the following form, 
large solar mixing angle can arise 
for some choice of the parameters in $\scriptstyle M_{\nu_{RR}}$, 
leading to LMA solution for solar neutrinos, 
\begin{equation}
\scriptstyle M_{\nu_{RR}} = \left(\begin{array}{ccc}
\scriptstyle c^{2}\eta^{2} & \scriptstyle -b\epsilon\eta & \scriptstyle a \eta\\
\scriptstyle -b\epsilon\eta & \scriptstyle \epsilon^{2} & \scriptstyle -\epsilon\\
\scriptstyle a \eta & \scriptstyle -\epsilon & \scriptstyle 1
\end{array}\right)\scriptstyle \cdot \Lambda_{R}, \quad
\scriptstyle M_{\nu}^{eff} = \scriptstyle \left(\begin{array}{ccc}
\scriptstyle 0 & \scriptstyle -\epsilon & \scriptstyle 0\\
\scriptstyle -\epsilon & \scriptstyle 0 & \scriptstyle 2\epsilon\\
\scriptstyle 0 & \scriptstyle 2\epsilon & \scriptstyle 1
\end{array}\scriptstyle \right)
m_{u}^{2}/\lambda_{R}.
\end{equation}
The value for $\scriptstyle |U_{e3}|$ is predicted to be small.
 
{\bf Large mixing from RGE's:} This class of models have been considered, for example, in  
\cite{Mohapatra:2003tw}. 
The RG evolution of the effective neutrino mass matrix is given by,
$\scriptstyle
dm_{\nu}/dt = - \{ \kappa_{u} m_{\nu} + m_{\nu} P + P^{T} m_{\nu} \}$
where
$
{\scriptstyle
P \simeq -\frac{1}{32\pi^{2}}\frac{h_{\tau}^{2}}{\cos^{2}\beta}diag(0,0,1), \quad
\kappa_{u} \simeq \frac{1}{16\pi^{2}}[\frac{6}{5}g_{1}^{2} + 6 g_{2}^{2} 
-6 \frac{h_{t}^{2}}{\sin^{2}\beta}].
}$
If one assumes nearly degenerate mass pattern and same Majorana CP phases at the 
GUT scale, the parameters $s_{12}$ and $s_{23}$ are driven to be large, 
while corrections to $m_{i}$ are small. 
Assuming leptonic mixing matrix is identical to the CKM matrix at the GUT scale.  
Starting with $\scriptstyle s_{12}^{0} \simeq \lambda$, $\scriptstyle s_{23}^{0} \simeq 
\mathcal{O}(\lambda^{2})$, and $\scriptstyle s_{13}^{0} \simeq \mathcal{O}(\lambda^{3})$, 
one obtains $\scriptstyle \sin^{2}2\theta_{atm}=0.99, \; 
\sin^{2}2\theta_{\odot}=0.87, \; \sin\theta_{13}=0.08$ at the weak scale.
These GUT scale conditions can be understood in Type II seesaw mechanism, with 
$\scriptstyle M_{\nu_{LL}}$ term (due to the coupling to an SU(2) triplet Higgs) 
dominates: because $\scriptstyle M_{\nu_{LL}} \sim I \cdot m_{\nu_{LL}}$ 
dominates, one obtains the nearly degenerate mass spectrum; as the flavor mixing 
is due to the usual seesaw term, the mixing angle of the 
resulting mass matrix is CKM-like. Thus the two conditions needed for enhancing the mixing 
angles are satisfied. The $\scriptstyle U_{e3}$ element is also amplified by the RG flow, 
with the low energy prediction close to the sensitivity of current experiments.

{\bf Large mixing from $b-\tau$ unification:} This class of models have been 
considered, for example, in Ref.\cite{Bajc:2002iw,Goh:2003sy}. 
It has a minimal Higgs sector which contains $\{10, 126, 45, 54 \}$.
The following mass relations arise due to SO(10) symmetry,
$\scriptstyle{M_{u}} \scriptstyle{=} \scriptstyle{f <10> + h <\overline{126}>}
, \,
\scriptstyle{M_{d}} \scriptstyle{=} \scriptstyle{f <10> + h <\overline{126}>}
, \,
\scriptstyle{M_{e}} \scriptstyle{=} \scriptstyle{f <10> -3 h <\overline{126}>}
, \,
\scriptstyle{M_{\nu_{LR}}} \scriptstyle{=} \scriptstyle{f <10> -3 h <\overline{126}>}$.
As there are only one 10 and one 126 Higgs representations, all mass terms are governed 
by two Yukawa matrices, $f$ and $h$.  
The small neutrino masses are explained by the Type II see-saw 
mechanism with the assumption that the LH Majorana mass 
term dominates over the usual Type I see-saw term, 
$\scriptstyle M_{\nu}^{eff} = M_{\nu_{LL}} - M_{\nu_{LR}}M_{\nu_{RR}}^{-1}M_{\nu_{LR}}^{T}$. 
The mass terms $\scriptstyle M_{\nu_{LL}}$ and $\scriptstyle M_{\nu_{RR}}$ are both due 
to the coupling to $\scriptstyle \overline{126}$, leading to 
$\scriptstyle M_{\nu_{LL}} \sim h v_{ew}^{2}/v_{R}$ and 
$\scriptstyle M_{\nu_{RR}} \sim h v_{R}$.
In this minimal scheme, we have the following sum-rule
$\scriptstyle M_{\nu}^{eff} = c (M_{d}-M_{e})$. 
The down-type quark and charged lepton mass matrices can be parameterized in 
terms of Wolfenstein parameter as
\begin{equation}
\scriptstyle M_{b,\tau} \sim
\left(
\begin{array}{ccc}
\scriptstyle{\lambda^{3}} & \scriptstyle{\lambda^{3}} & \scriptstyle{\lambda^{3}}\\
\scriptstyle{\lambda^{3}} & \scriptstyle{\lambda^{2}} & \scriptstyle{\lambda^{2}}\\
\scriptstyle{\lambda^{3}} & \scriptstyle{\lambda^{2}} & \scriptstyle{1}
\end{array}\scriptstyle 
\right) m_{b,\tau}
\end{equation}
For some value of $\scriptstyle \tan\beta$ (small values are preferred), 
the deviation from $b-\tau$ unification at the GUT scale is
$\scriptstyle m_{b}(M_{GUT} )- m_{\tau}(M_{GUT}) \simeq \mathcal{O}(\lambda^{2}) m_{\tau}$
which leads to a bi-large mixing pattern in $M_{\nu}$. Generic predictions of this model are 
$\scriptstyle \sin^{2}2\theta_{23}<0.9$ and $\scriptstyle \sin^{2}2\theta_{12}>0.9$, 
making the model testable. The prediction of this model for the value of 
$\scriptstyle |U_{e3}|$ is large. 
This is a consequence of the atmospheric mixing angle being maximal.
 
{\bf Comparison of models:} 
Predictions of selected models for $\scriptstyle \sin\theta_{13}$ are summarized in 
Table 1.
\begin{table}{\footnotesize
\begin{tabular}{lccr}\hline
Model &  family symmetry &  solar solution & $\sin\theta_{13}$\\
\hline\\
Albright-Barr\cite{Albright:2001uh} & $U(1)$ & LMA & 0.014\\
Babu-Pati-Wilczek\cite{Babu:1998wi} & $U(1)$ & SMA & $5.5 \times 10^{-4}$\\
Blazek-Raby-Tobe\cite{Blazek:1999hz} & $U(2) \times U(1)^{n}$ & LMA & 0.049\\
Berezhiani-Rossi\cite{Berezhiani:2000cg} & $SU(3)$ & SMA & $\mathcal{O}(10^{-2})$ 
\\
Chen-Mahanthappa\cite{Chen:2002pa} & $SU(2)$ & LMA & 0.149\\
Kitano-Mimura\cite{Kitano:2000xk}& $SU(3) \times U(1)$ & LMA 
&  $\sim \lambda \sim 0.22$ \\
Maekawa\cite{Maekawa:2001uk} & $U(1)$ & LMA & $\sim \lambda \sim 0.22$ \\
Raby\cite{Raby:2003ay} & $3 \times 2$ seesaw with $SU(2)_{F}$ 
& LMA & $\sim m_{\nu_{2}}/2m_{\nu_{3}} \sim \mathcal{O}(0.1)$\\
Ross-Velasco-Sevilla\cite{Ross:2002fb} & $SU(3)$ & LMA & 0.07\\
\hline
\\
Frampton-Glashow\cite{Frampton:2002qc} & $3 \times 2$ seesaw &
LMA  & $\sim m_{\nu_{2}}/2m_{\nu_{3}} \sim \mathcal{O}(0.1)$ \\
$\quad$ -Yanagida &&&\\
Mohapatra-Parida\cite{Mohapatra:2003tw} & RG enhancement &
LMA & $0.08-0.10$\\
$\quad$ -Rajasekeran &&&\\
\hline
\end{tabular}
\caption{\footnotesize Predictions for $\sin\theta_{13}$ of various models. 
The upper bound from CHOOZ experiment is $\sin\theta_{13} < 0.24$. 
First nine models use $SO(10)$. Last two models are not based on $SO(10)$.}
\label{theta13} }
\end{table}
A general observation is the following:
(i) large $\scriptstyle U_{e3} \sim \mathcal{O}(0.1)$ 
(can be probed by conventional/superbeam) 
arises in models with symmetric texture, in 
models based on anarchy, in 
models with RG enhanced leptonic mixing, and in  
minimal models with approximate $\scriptstyle b-\tau$ unification;
(ii) intermediate $U_{e3}$ value $\scriptstyle \sim (0.05-0.07)$ 
(can be probed by superbeam) arises in models with asymmetric texture; 
(iii) small $\scriptstyle U_{e3}$ (neutrino factory may be needed) arises in  
models with lop-sided texture. The mass hierarchy can be probed by 
long baseline experiments with distance $\scriptstyle > 900 Km$, 
which can be used to distinguish different models. 
A general observation is
(i) normal hierarchy arises in 
SO(10) models with $\scriptstyle SU(3)_{H}$, $\scriptstyle SU(2)_{H}$, 
$\scriptstyle U(1)_{H}$, in  
minimal SO(10) models with $\scriptstyle b-\tau$ unification, and in 
SO(10) models with $\scriptstyle 3\times 2$ see-saw; 
(ii) inverted hierarchy arises in 
models with $\scriptstyle L_{e}-L_{\mu}-L_{\tau}$ horizontal symmetry; 
(iii) nearly degenerate arises in 
SO(10) models with RG enhanced lepton mixing and in models with anarchy.
This work was supported by US DOE under Grant No. 
DE-AC02-98CH10886 and DE-FG03-95ER40894.

%% If the bibliography is produced without BibTeX comment out the

\footnotesize{

\bibliography{sample}

\IfFileExists{\jobname.bbl}{}
 {\typeout{}
  \typeout{******************************************}
  \typeout{** Please run "bibtex \jobname" to optain}
  \typeout{** the bibliography and then re-run LaTeX}
  \typeout{** twice to fix the references!}
  \typeout{******************************************}
  \typeout{}
 }

\end{document}